\documentstyle[a4,aps,epsf]{revtex}

\def\s{\sigma}

\def\up{\uparrow}
\def\dd{\downarrow}
\def\las{\langle}
\def\ras{\rangle}

\def\nn{\nonumber}

\def\kp{{\bf k_\parallel}}

\begin{document}
\title{Spin-filter magnetoresistance in magnetic barrier junctions}
\author{ Alireza Saffarzadeh\thanks{E-mail: a-saffar@tehran.pnu.ac.ir}\\
Department of Physics, Tehran Payame Noor University, Fallahpour St.,\\
Nejatollahi St., Tehran, Iran}
\date{\today}
\maketitle

\begin{abstract}
The tunnel current and magnetoresistance (TMR) are investigated in magnetic
tunnel junctions consisting of a spin-filter tunnel barrier, sandwiched between
a ferromagnetic (FM) electrode and a nonmagnetic (NM) electrode. The
investigations are based on the transfer matrix method and the free-electron
approximation. The numerical results show that the spin transport depends on
the relative magnetization orientation of the FM electrode and the
spin-filter barrier, such that the tunnel current reaches its maximum when the
magnetic moments of the FM electrode and the magnetic barrier are parallel.
It is also found that the TMR increases with increasing the applied voltage.
\end{abstract}

\newpage
\section{\bf Introduction}
Since the observation of tunnel magnetoresistance (TMR) in magnetic tunnel
junctions \cite{Jul}, there has been a lot of interest in structures consisting
of two ferromagnetic (FM) electrodes separated by one insulator or
semiconductor (FM/I(S)/FM). This is due to the possible applications as
magnetic field sensors and memory cells in magnetic random access memories
\cite{Mo,Parkin}. The TMR in the magnetic tunnel junctions is limited by the
spin polarizations of the FM electrodes. Recently, large magnetoresistance
ratios as high as 30\% were reported in Fe/Al$_2$O$_3$/Fe \cite{Miy} and
CoFe/Al$_2$O$_3$/Co \cite{Moo}.

The TMR of such FM/I(S)/FM tunnel junctions can be understood in terms of a
two-band model in which the energy bands of the FM electrodes are
split into spin-up and spin-down bands with different density of states at the
Fermi energy. When the magnetization of the electrodes is parallel, the
spin-up (spin-down) electrons tunnel from a majority (minority) spin to a
majority (minority) spin band, whereas in the antiparallel alignment, the
spin-up (spin-down) electrons are forced to tunnel from a majority (minority)
spin to a minority (majority) spin band. This gives rise to a change in the
conductance when the magnetizations are switched. During the last 10 years,
a lot of theoretical papers were published on this topic (see reviews in Refs.
\cite{Mes,Zhang}).

Using magnetic insulating layers, instead of having FM electrodes, one can
obtain very high values for the TMR \cite{DCW,Lec,Saff}. Tunneling experiments
using ferromagnetic semiconductors (FMSs), in particular the Eu chalcogenides,
as the magnetic tunnel barriers \cite{Mood1,Hao}, have shown spin polarization
in the tunnel current which in favorable cases exceeds 99\% \cite{Mood2}.
When a magnetic barrier, which acts as spin filter, is used in a tunnel
junction, due to the spin splitting of its conduction band below $T_C$ (Curie
temperature), tunneling electrons see a spin dependent barrier height.
In this case the probability of tunneling for one spin channel will be
much larger than the other, and a highly spin
polarized current may result. More recently, LeClair et al.
\cite{Lec} using a spin-filter barrier and an FM electrode, obtained a large
TMR in an Al/EuS/Gd tunnel junction which is a new method for injecting
spins into semiconductors. The spin filtering and the magnetoresistance have
also been investigated theoretically in semimagnetic semiconductors
\cite{Egues}.

In this paper, using the transfer matrix method, we study theoretically the
effect of the thickness of the magnetic barrier and the applied bias
on the TMR for tunneling through a FM/FMS/NM tunnel junction with an
asymmetrical barrier.
We assume that the electron wave vector parallel to the interfaces and the
spin direction of the electron are conserved in the tunneling process through
the whole system.

In section 2, we describe the model and present a general formula for the
tunneling current through the magnetic tunnel junction. Numerical results for
a typical tunnel junction are presented in section 3. A brief summary is given
in section 4.

\section{\bf Model and formalism}

Consider a FM/FMS/NM sandwiched structure in the presence of DC bias $V_a$ as
shown in Fig. 1, where $\phi_L$ and $\phi_R$ are the barrier heights in the
left- and right-hand side of the FMS layer above $T_C$, respectively. For a
tunnel junction
with different electrode materials, the difference in barrier height at both
metal-insulator interfaces makes the barrier oblique, resulting in an
asymmetric current-voltage behavior. In this case, an asymmetry parameter
$\Delta\phi=\phi_L-\phi_R$ is introduced at zero bias to account for the
tilted barrier potential \cite{Simmons}. Here we consider the case that the
tunneling electron with energy $E_x$ is incident from the left and transmits
to the right along the $x$ direction. In a free-electron approximation of the
spin-polarized conduction electrons, the longitudinal part of the effective
one-electron Hamiltonian may be written as
\begin{equation}\label{H}
H_x=-\frac{\hbar^2}{2m_j^*}\frac{d^2}{dx^2}+U_j(x)+V^\s-{\bf h}(x)\cdot{\bf\s},
\end{equation}
where $m^*_j$ ($j$=1-3) are the electron effective masses in three regions
labeled in Fig. 1, and
\begin{equation}\label{F}
U_j(x)=\left\{\begin{array}{cc}
0, & x<0 \ ,\\
E_{FL}+\phi_L-(\Delta\phi+eV_a)x/d, & 0\leq x<d\  ,\\
-eV_a, & x\geq d\  ,
\end{array}\right.
\end{equation}
where $E_{FL}$ is the Fermi energy of the FM electrode and $d$ is the barrier
width. $V^{\s}$ which is a spin-dependent potential, denotes the $s-f$
exchange coupling between the spin of tunneling electrons and the localized
$f$ spins in the magnetic barrier. This term, within the mean field
approximation, is proportional to the thermal average of the $f$ spins,
$\las S_z\ras$ (a 7/2 Brillouin function), and can be written as
$V^{\s}=-I\s\las S_z\ras$. Here, $\s=\pm 1$ which corresponds to
$\s=\up,\dd$, respectively and $I$ is the $s-f$ exchange constant in the
FMS layer. $-{\bf h}(x)\cdot{\bf\s}$ is the internal exchange energy where
${\bf h}(x)$ is the molecular field in the FM electrode and $|{\bf h}|=h_0$.
Although the transverse momentum $\hbar\kp$ is omitted from
the above notations, the summation over $\kp$ is carried out in our
calculations.

The Schr\"odinger equation for a biased barrier layer can be simplified by a
coordinate transformation whose solution is the linear combination of the Airy
function Ai[$\rho(x)$] and its complement Bi[$\rho(x)$] \cite{Abram}.
Considering all three regions of the FM/FMS/NM junction shown in Fig. 1, the
eigenfunctions of the Hamiltonian (\ref{H}) with eigenvalue $E_x$ have the
following forms
\begin{equation}\label{psi}
\psi_{j\s}(x)=\left\{\begin{array}{cc}
A_{1\s}e^{ik_{1\s}x}+B_{1\s}e^{-ik_{1\s}x}, & x<0 \ ,\\
A_{2\s}\mbox{Ai}[\rho_\s(x)]+B_{2\s}\mbox{Bi}[\rho_\s(x)], & 0\leq x<d\  ,\\
A_{3\s}e^{ik_{3}x}+B_{3\s}e^{-ik_{3}x}, & x\geq d\  ,
\end{array}\right.
\end{equation}
where
\begin{equation}
k_{1\s}=\sqrt{2m_1^*(E_x+h_0\s)}/\hbar \  ,
\end{equation}
and
\begin{equation}
k_{3}=\sqrt{2m_3^*(E_x+eV_a+E_{FR}-E_{FL})}/\hbar\  ,
\end{equation}
are the electron wave vectors along the $x$ axis. $A_{j\s}$ and $B_{j\s}$ are
constants to be determined from the boundary conditions,
while
\begin{equation}
\rho_\s(x)=\frac{x}{\lambda_0}+\beta_\s\  ,
\end{equation}
with
\begin{equation}
\lambda_0=-\left[\frac{\hbar^2d}{2m^*_2(\Delta\phi+eV_a)}\right]^{1/3}\  ,
\end{equation}
\begin{equation}
\beta_\s=\frac{[E_x-E_{FL}-\phi_L-V^\s]d}{(\Delta\phi+eV_a)
\lambda_0}\  .
\end{equation}

Upon applying the boundary conditions such that the wave functions and their
first derivatives are matched at each interface point $x_j$, i.e.,
$\psi_{j,\s}(x_j)=\psi_{j+1,\s}(x_j)$ and $(m^*_j)^{-1}[d\psi_{j,\s}(x_j)/dx]=
(m^*_{j+1})^{-1}[d\psi_{j+1,\s}(x_j)/dx]$, we obtain a matrix formula that
connects the coefficients $A_{1\s}$ and $B_{1\s}$ with the coefficients
$A_{3\s}$ and $B_{3\s}$ as follows:
\begin{eqnarray}
\left[\begin{array}{c}A_{1\s}\\B_{1\s}
\end{array}\right]
=M_{total}\left[\begin{array}{c}A_{3\s}\\B_{3\s}\end{array}\right]\  ,
\end{eqnarray}
where
\begin{eqnarray}\label{M}
M_{total}&=&\frac{k_{3}}{k_{1\s}}
\left[\begin{array}{cc}
ik_{1\s}& \frac{1}{\lambda_0}\frac{m_1^*}{m_2^*}\\
ik_{1\s}&-\frac{1}{\lambda_0}\frac{m_1^*}{m_2^*}
\end{array}\right]
\left[\begin{array}{cc}
\mbox{Ai}[\rho_\s(x=0)]&\mbox{Bi}[\rho_\s(x=0)]\\
\mbox{Ai}'[\rho_\s(x=0)]&\mbox{Bi}'[\rho_\s(x=0)]
\end{array}\right]\nn\\&&
\times\left[\begin{array}{cc}
\mbox{Ai}[\rho_\s(x=d)]&\mbox{Bi}[\rho_\s(x=d)]\\
\mbox{Ai}'[\rho_\s(x=d)]&\mbox{Bi}'[\rho_\s(x=d)]
\end{array}\right]^{-1}
\left[\begin{array}{cc}
ik_{3}& \frac{1}{\lambda_0}\frac{m_3^*}{m_2^*}\\
ik_{3}&-\frac{1}{\lambda_0}\frac{m_3^*}{m_2^*}
\end{array}\right]^{-1}\nn\\&&
\times\left[\begin{array}{cc}
e^{-ik_{3}d}&0\\
0&e^{ik_{3}d}
\end{array}\right]^{-1}\  .
\end{eqnarray}
Since there is no reflection in region 3, the coefficient $B_{3\s}$ in
Eq. (\ref{psi}) is zero and the transmission coefficient of the spin $\s$
electron which is defined as the ratio of the transmitted flux to the incident
flux can be written as
\begin{equation}\label{Ps}
T_\s(E_x,V_a)=\frac{k_{3}m_1^*}{k_{1\s}m_3^*}\left|\frac{1}{M_{total}^{11}}
\right|^2\  ,
\end{equation}
where $M_{total}^{11}$ is the left-upper element of the matrix $M_{total}$
defined in Eq. (\ref{M}).

At $T=0$ K, the spin-dependent current density for the magnetic tunnel
junction in the free-electron model is given by the formula \cite{Duke}
\begin{equation}\label{J}
J_\s=\frac{em^*_1}{4\pi^2\hbar^3}\left[
eV_a\int_{E_0^\s}^{E_F-eV_a}T_\s(E_x,V_a)dE_x+
\int_{E_F-eV_a}^{E_F}(E_F-E_x)T_\s(E_x,V_a)dE_x\right]\  ,
\end{equation}
where $E_0^\s$ is the lowest possible energy that will allow transmission and
is given by $E_0^\up=$max$\{-h_0,-(eV_a+E_{FR}-E_{FL})\}$ for spin-up and
$E_0^\dd=h_0$ for spin-down electrons.

The tunnel conductance per unit area is given by $G=\sum_{\s}J_{\s}/V_a$. In
this case, the TMR can be described quantitatively by the relative
conductance change as
\begin{equation}\label{Tmr}
\mbox{TMR}=\frac{G_{\up\up}-G_{\up\dd}}{G_{\up\up}},
\end{equation}
where $G_{\up\up}$ and $G_{\up\dd}$ correspond to the conductances in the
parallel and antiparallel alignments of the magnetizations, respectively.

\section{\bf Numerical results}
Taking the Fe/EuS/Al tunnel junction as an example, we calculate the tunnel
current and TMR of the junction according to the Eqs. (\ref{Ps})-(\ref{Tmr}).
We have chosen Fe and EuS because they have cubic structures and the lattice
mismatch is only 4\% \cite{Dem}. The appropriate parameters for EuS which
have been used in this article are: $S$=7/2, $I$=0.1 eV \cite{Nolting}.
The parameters $E_{FL}$ and $h_0$ for Fe layer are
taken corresponding to $k_{F\up}$=1.09 \AA$^{-1}$ and
$k_{F\dd}$=0.42 \AA$^{-1}$ (for itinerant $d$ electrons) \cite{Stear}.
In the Al layer $E_{FR}=11.7$ eV
\cite{Ash}. The barrier heights at the interfaces, which can be derived from
the work functions \cite{Park} of Fe and Al and the electron affinity of EuS
\cite{Mood1}, are taken as $\phi_L$=1.94 eV and $\phi_R$=1.7 eV.
In practice, the effective masses of electrons may differ
from that of free electron, but here for simplicity, we assume all electrons
have the same mass, $m$ as free electrons. We show the numerical results at
$T$=0 K. Thus, in the case of parallel (antiparallel) alignment
$\las S_z\ras=S$ ($\las S_z\ras=-S$).

In Fig. 2, the TMR is shown as a function of the width of the magnetic
barrier $d$, when the bias voltage $V_a$=0.5 V is applied to the junction.
At zero temperature and for the nearly normal incidence, the electrons with
$E_x$ near $E_F$ carry most of the current. Furthermore, the Fermi energy
and the barrier heights at fixed temperature, are constant. Therefore, with
increasing the thickness of the EuS layer from zero, when the thickness
approaches half the wavelength of the electron wave in the
magnetic barrier, the tunnel conductance with the parallel alignment increases
faster than with the antiparallel alignment. In this case, we can expect a peak
in the TMR. By increasing the thickness further, the tunnel conductance and
hence the TMR decreases exponentially.

Figure 3 shows the spin-dependent current densities for spin-up and spin-down
electrons as a function of applied bias.
In the parallel alignment, the tunnel current for spin-up electrons is much
higher than the spin-down ones, while in the antiparallel alignment, the
discrepancy between the tunnel currents of the two spin channels increases very
slowly. The origin of this effect is that the tunnel current for each spin
channel depends on the density of states in the Fe layer and the heights of
the tunnel barrier. In the parallel (antiparallel) alignment, the majority
spin electrons tunnel through a barrier with low (high) height, whereas the
minority spin electrons tunnel through a barrier with high (low) height, thus
we can expect a high (low) spin-polarized current.

It is clearly observed that for both the parallel and antiparallel
alignments at low voltages the tunnel current curves vary linearly.
With increasing the bias voltage, the effective width of the barrier becomes
narrower and hence the slope of the curves increases. This effect is very
strong for spin-up electrons in the parallel alignment, and causes the TMR
increases with increasing the bias voltage.

In order to reveal another aspect of the spin filtering phenomenon in the
system, we have displayed the voltage dependence of the TMR in Fig. 4. As the
figure shows for $d$=7 and 10 \AA, with increasing the bias voltage the TMR
increases linearly and decreases at high voltages. This decrease which is
not identical for different barrier thicknesses is a consequence of the
narrowing of the barrier width. At low voltages, the tunneling probability
for one spin channel is higher than the other, whereas at high voltages,
by narrowing the barrier, this tunneling probability increases for both
channels; thus, the tunnel conductance ratio, $G_{\up\dd}/G_{\up\up}$, tends
toward unity, and hence the TMR decreases. This dependence of the TMR on bias
voltage is not similar to the results of the FM/I(S)/FM tunnel junctions
\cite{Dav}, because in the present study, we have a spin-dependent barrier.
In FM/I(S)/FM junctions, both the majority and minority electrons see identical
barrier heights. In this case, as the bias voltage increases, the imbalance
between the number of majority and minority tunneling states diminishes and,
therefore, the TMR decreases.

From these results one can see that, for the FM/FMS/NM tunnel junctions,
the spin transport and hence the TMR can be controlled by the applied bias
and the thickness of the FMS layer.

\section{\bf Concluding remarks}
In summary, we have shown how using a magnetic barrier and a FM electrode
one leads to a new magnetoresistive device. The obtained results for
spin-filter magnetoresistance show that the spin currents depends on the
relative magnetization orientation of the FM electrode and the spin-filter
barrier. It was also found that the low spin-polarized current will be
amplified through the magnetic barrier; thus, a very large spin-polarized
current from the FM electrode is injected into the NM electrode and a high
value for the TMR is then obtained.

Our analysis of the TMR can be potentially useful to achieve larger TMR by
optimally adjusting the material parameters. The two materials systems,
ferrites and garnets \cite{Handl} are both insulating and ferrimagnetic above
room temperature. Therefore, they are well suited for obtaining large TMR in
FM/FMS/NM. However, for successful spintronics applications, future efforts
will have to concentrate on fabricating FM semiconductors in which
ferromagnetism will persist at higher temperatures.

\section*{Acknowledgments}
I would like to express sincere thanks to Dr. Patrick R. LeClair for helpful
discussion.

\newpage
\begin{figure}
\begin{center}
\leavevmode\hbox{\epsfxsize=0.9\textwidth\epsffile{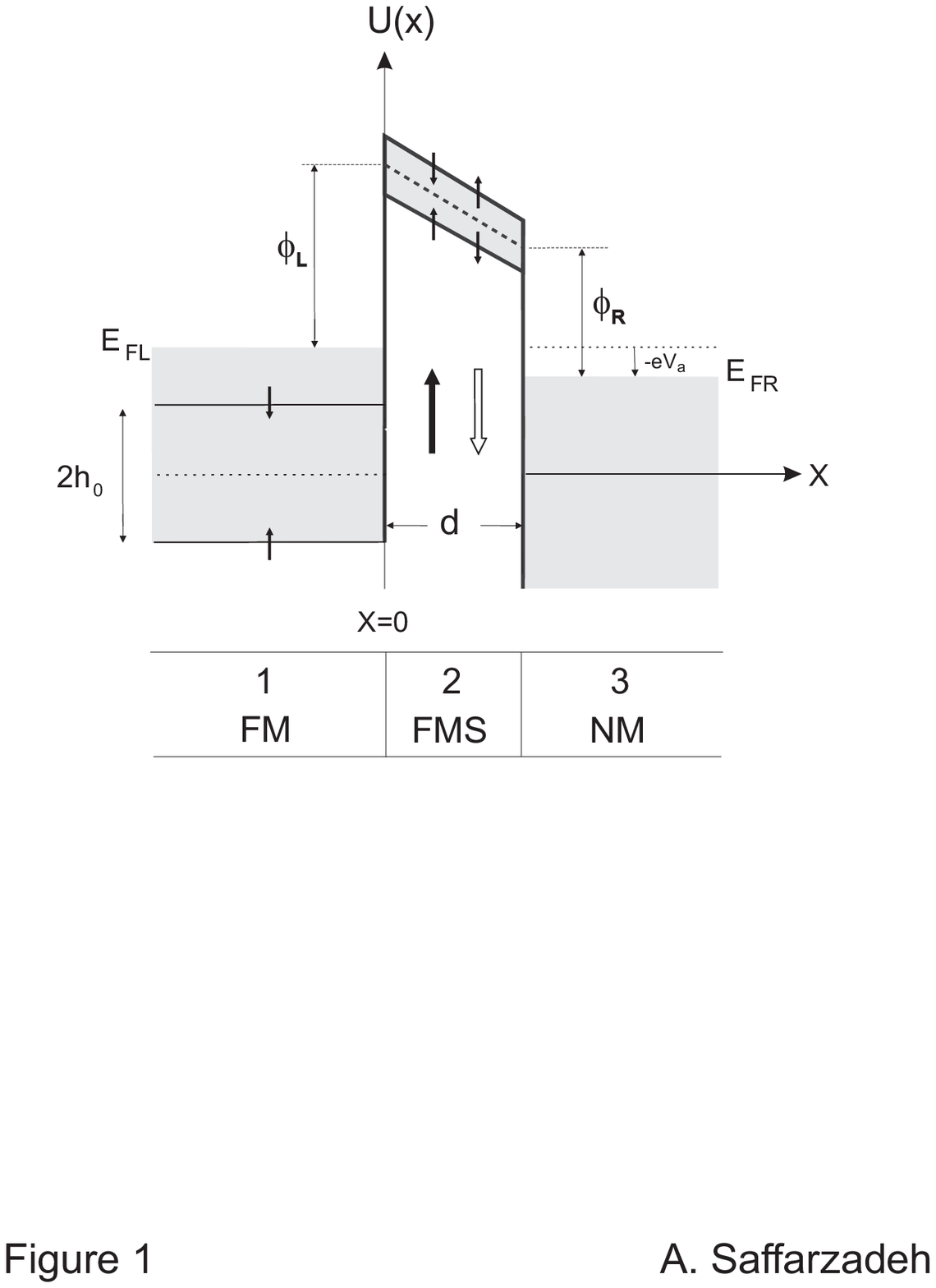}}
\end{center}
\caption{Spin-dependent potential profile for FM/FMS/NM tunnel junctions in
the presence of a positive bias $V_a$. In the FMS layer, the dashed line
represents the bottom of the conduction band at $T\geq T_c$ and the thin arrows
indicate the bottom of the conduction band for spin-up and spin-down
electrons at $T<T_C$. The direction of magnetization in the FM layer is fixed in the
$+z$, while the magnetization in the magnetic barrier is free to be flipped
into either the $+z$ or $-z$ direction, as indicated by thick arrows. The zero
of energy is taken at the middle of bottoms for majority-spin band and
minority-spin one in the FM layer.}
\end{figure}

\begin{figure}
\begin{center}
\leavevmode\hbox{\epsfxsize=0.9\textwidth\epsffile{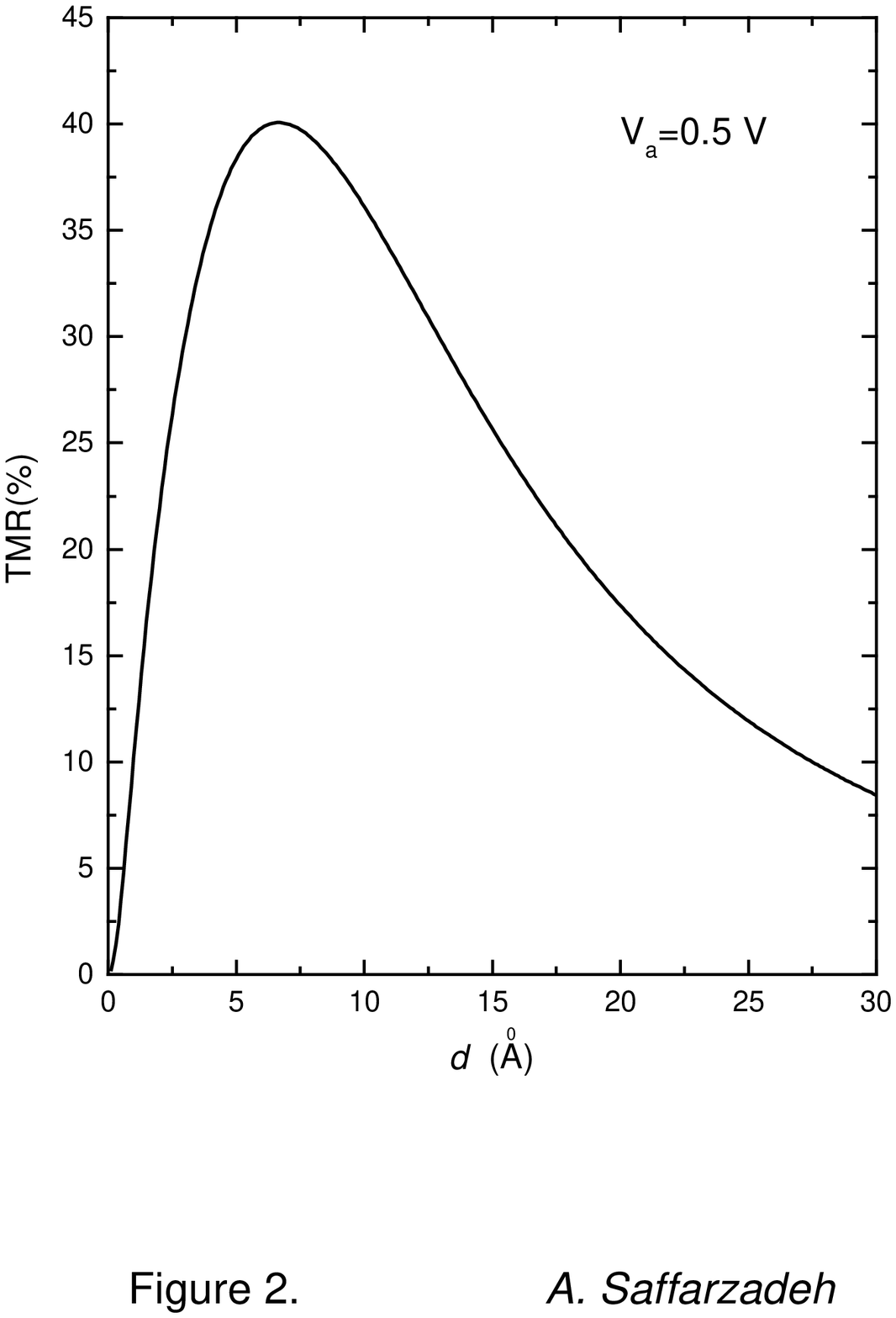}}
\end{center}
\caption{Dependence of the TMR on the thickness of EuS layer.}
\end{figure}

\begin{figure}
\begin{center}
\leavevmode\hbox{\epsfxsize=0.9\textwidth\epsffile{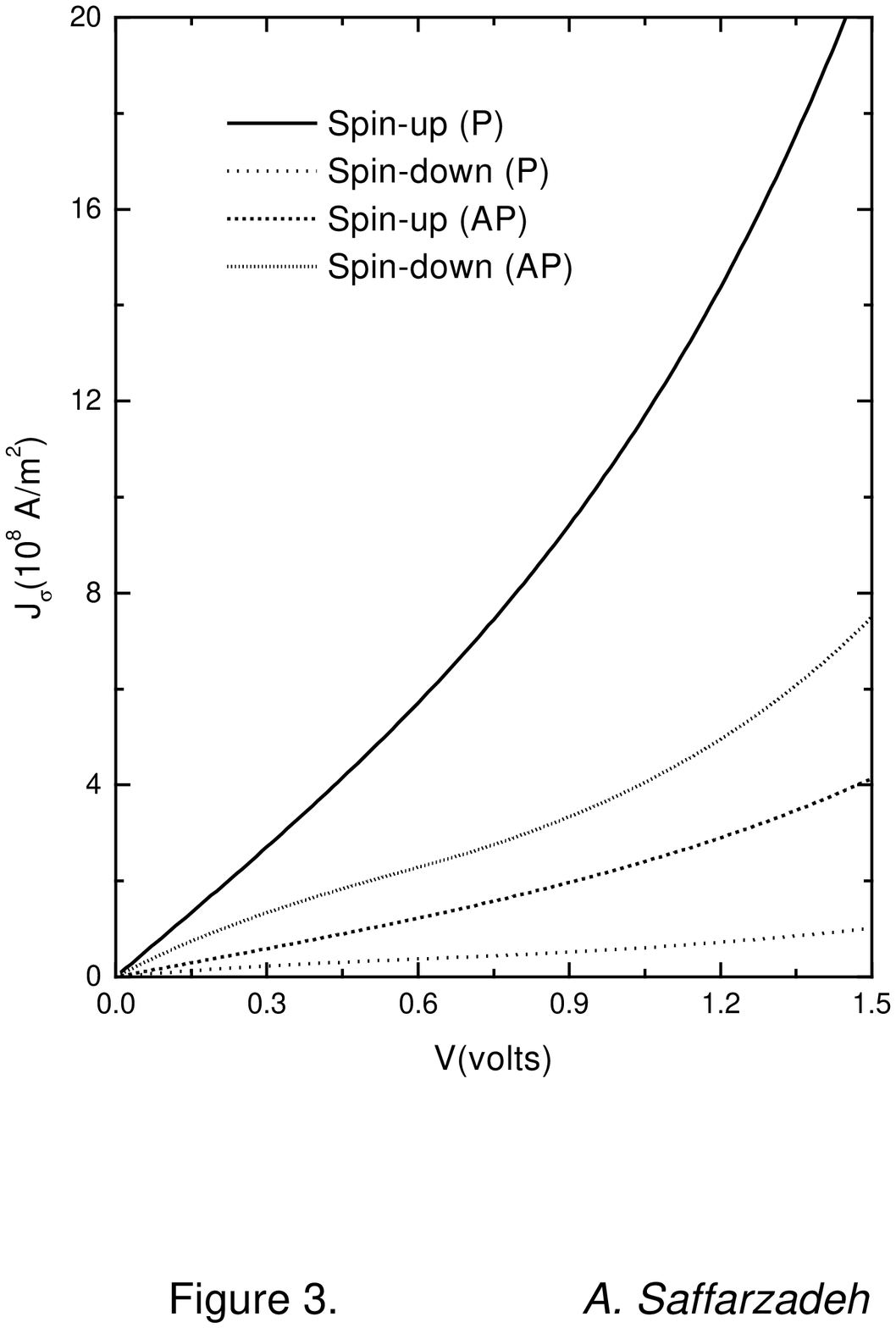}}
\end{center}
\caption{Dependence of the current densities $J_\s$ on bias voltage calculated
for $d$=7 \AA\ in the parallel (P) and antiparallel alignments (AP).}
\end{figure}

\begin{figure}
\begin{center}
\leavevmode\hbox{\epsfxsize=0.9\textwidth\epsffile{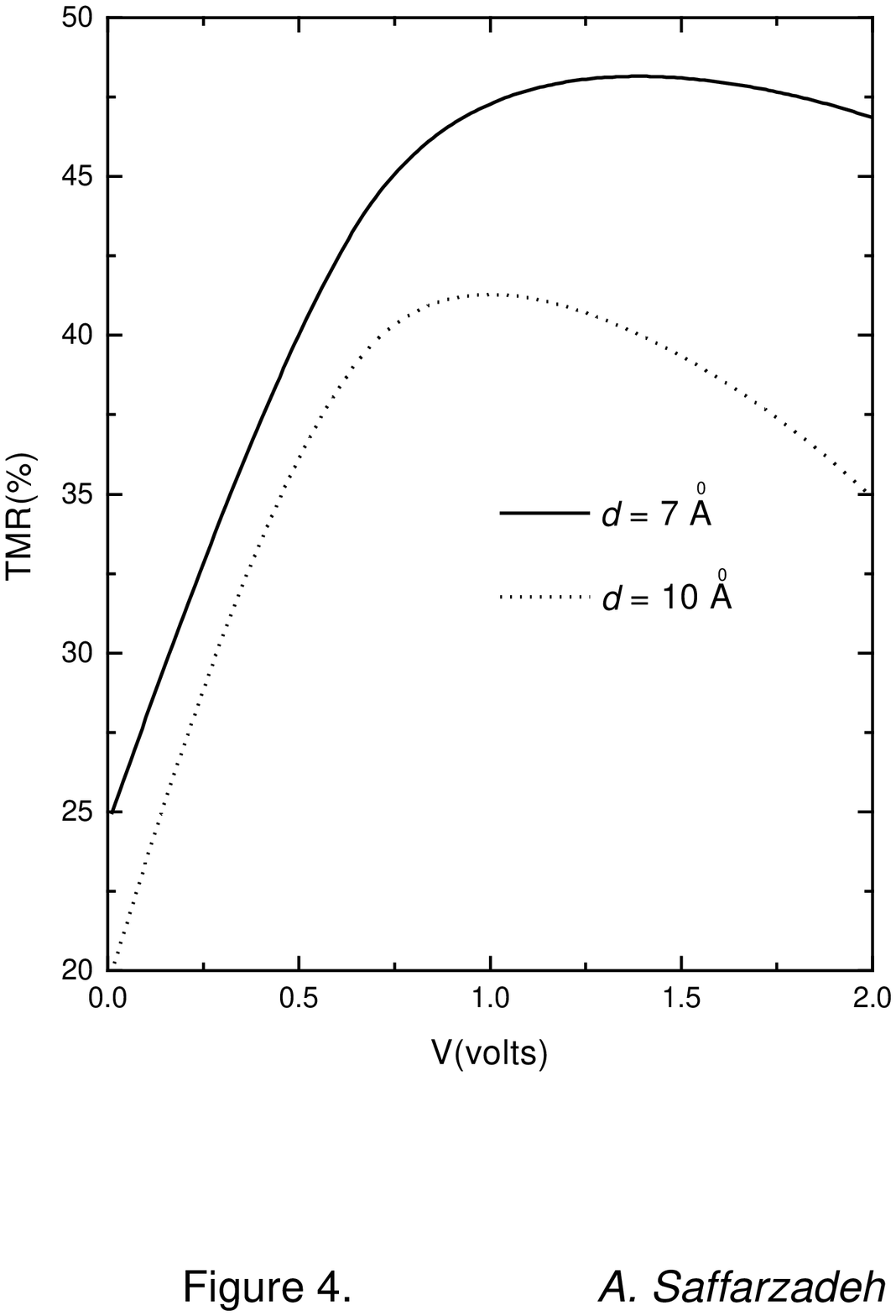}}
\end{center}
\caption{Dependence of the TMR on bias voltage calculated for $d$=7 and
10 \AA.}
\end{figure}

\end{document}